\documentclass[aps,pre,floatfix,footinbib,twocolumn,showpacs]{revtex4}

\usepackage{graphicx}
\usepackage{amsmath}
\usepackage{amssymb}

\begin{document}
\title{Diffusing proteins on a fluctuating membrane: Analytical theory and simulations}
\author{Ellen Reister-Gottfried}
\author{Stefan M.~Leitenberger}
\author{Udo Seifert}
\affiliation{Universit\"at Stuttgart, II. Institut f\"ur Theoretische Physik, D-$70550$  Stuttgart, Germany.}

\date{\today}

\begin{abstract}
  Using analytical calculations and computer simulations we consider both the
  lateral diffusion of a membrane protein and the fluctuation spectrum of the
  membrane in which the protein is embedded. The membrane protein interacts
  with the membrane shape through its spontaneous curvature and bending
  rigidity. The lateral motion of the protein may be viewed as diffusion in an
  effective potential, hence, the effective mobility is always reduced compared
  to the case of free diffusion. Using a rigorous path-integral approach we
  derive an analytical expression for the effective diffusion coefficient for
  small ratios of temperature and bending rigidity, which is the biologically
  relevant limit. Simulations show very good quantitative agreement with our
  analytical result. The analysis of the correlation functions contributing to
  the diffusion coefficient shows that the correlations between the stochastic
  force of the protein and the response in the membrane shape are responsible
  for the reduction.

  Our quantitative analysis of the membrane height correlation spectrum shows
  an influence of the protein-membrane interaction causing a
  distinctly altered wave-vector dependence compared to a free
  membrane. Furthermore, the time correlations exhibit the two relevant
  timescales of the system: that of membrane fluctuations and that of lateral
  protein diffusion with the latter typically much longer than the former. We
  argue that the analysis of the long-time decay of membrane height
  correlations can thus provide a new means to determine the effective
  diffusion coefficient of proteins in the membrane.
\end{abstract}
\maketitle
\section{Introduction}
Biomembranes are ubiquitous in life, mainly providing spatial
compartmentalisation. However, a membrane should not be viewed as a mere
barrier between different compartments, but serves as a place where a whole
variety of functions may take place, like ion or protein transport, signal
transduction, etc.~\cite{Alberts:1994}. These functions come about through
proteins that move along the membrane. From a physical perspective the lateral
diffusion of the proteins and the shape changes of the membrane caused upon
insertion of proteins are among the most interesting issues of these systems.

The recent progress in experimental techniques to measure lateral diffusion
coefficients, like fluorescence correlation spectroscopy~\cite{Chiantia:2009},
single particle tracking~\cite{Lommerse:2004}, or fluorescence recovery after
photobleaching~\cite{Reits:2001}, has revealed that many of the functions
performed by proteins are crucially influenced by the diffusive behavior of
the proteins~\cite{Cairo:2008}. Apart from the obvious biological relevance
lateral protein diffusion is also very challenging from a theoretical
perspective: Compared to diffusion in the bulk there is a subtlety in the
hydrodynamic equations describing the mobility in a two dimensional fluid,
since the solution of the two-dimensional Navier-Stokes equation diverges. In
order to overcome this so-called Stokes' paradox~\cite{Lamb:1959}, Saffman and
Delbr\"uck~\cite{Saffman:1975} considered the mobility of a very thin, rigid
object in a narrow almost two-dimensional fluid layer that is surrounded on
both sides by a further liquid. This work has received a lot of attention
since it is relevant for lateral protein diffusion. While some experiments
support their result ~\cite{Peters:1982,Lee:2003,Cicuta:2007}, more recent
observations for proteins cannot be explained by their
theory~\cite{Gambin:2006,Guigas:2006}.

Another aspect that makes diffusion interesting, particularly in membranes, is
that the membrane itself is subject to thermal fluctuations; thus the shape of
the membrane is also constantly changing.  Methods to analyse shape
fluctuations of membranes include off-specular x-ray
scattering~\cite{Charitat:2008} and video
microscopy~\cite{Faucon:1989,Pecreaux:2004}. In the latter method the contour
of a vesicle is detected from optical microscopy records taken at successive
timesteps.  The changes in the contour provide information on the fluctuation
spectrum that is used to deduce effective bending rigidities or surface
tensions. In a very recent study Rodr\'{i}guez-Garc\'{i}a et
al.~\cite{Rodriguez:2009} identified the influence of the bilayer nature of a
membrane as theoretical calculations~\cite{Seifert:1993} have previously
predicted. The influence of the density of inclusions embedded in a lipid
membrane on the effective bending rigidity was studied by Vitkova et
al.~\cite{Vitkova:2006}. In this work the peptide alamethicin was used as the
inclusion.  Bassereau and co-workers have studied more complicated systems
consisting of membranes with inserted proteins and have been able to study the
altered fluctuation spectrum of a membrane upon activation of the inserted
bacteriorhodopsin proteins~\cite{Girard:2005,Faris:2009}.

Theoretically, a bare membrane is well described as a continuous
two-dimensional sheet with a bending rigidity and an effective surface
tension. This model has been very successful in explaining a whole variety of
experimentally observed membrane morphologies~\cite{Seifert:1997}. Likewise,
membrane shape fluctuations are well captured by this simple model as shown in
video microscopy experiments~\cite{Faucon:1989,Pecreaux:2004}. The insertion
of additional proteins in a membrane requires an extension of this simple
continuous model to include the local interacion of a protein with the
membrane. While the influence of thermal membrane fluctuations on the
interaction between inclusions has previously been considered in several
studies~\cite{Goulian:1993,Netz:1995}, the influence on lateral diffusion or
the altered membrane height correlations is much less studied.  In previous
work involving both analytical calculations and simulations we and others
analyzed the geometric effect of measuring the diffusion coefficient from the
projected path of the
protein~\cite{Reister:2005,Reister:2007,Naji:2007}. While these studies only
included free diffusion, the lateral diffusion of an inclusion that interacts
with the membrane shape is considered in recent
studies~\cite{Reister:2005,Leitenberger:2007a,Gov:2006,Shlomovitz:2008,Naji:2009}.
If the membrane shape fluctuations were not influenced by the protein, the
effective diffusion coefficient would be \emph{increased} compared to the free
diffusion
coefficient~\cite{Reister:2005,Leitenberger:2007a,Shlomovitz:2008}. However,
this simplifying assumption represents too severe an approximation. By
including the backaction of the protein on the membrane fluctuations 
Naji et al.~\cite{Naji:2009} showed that in equilibrium the effective lateral
diffusion coefficient is \emph{reduced} for which they were able to give an approximate
expression. Parts of the current work are complementary to their study.

On a more collective level, the interaction between membrane and embedded
proteins can cause morphological changes.  Leibler~\cite{Leibler:1986} studied
a model with a protein density field that induces a spontaneous curvature
capable of causing an instability of the membrane. He finds two characteristic
time scales for membrane fluctuations, with one of them potentially unstable
in a certain wave vector range. Related work by Bivas and
M\'{e}l\'{e}ard~\cite{Bivas:2003} on bending elasticity and fluctuations of
spherical bilayer vesicles with additives reveals an additional characteristic
time scales. In their work the bilayer membrane comprises two individual
sheets such that the third timescale results from the friction between the two
layers.

Divet et al.~\cite{Divet:2002} studied an extension of Leibler's work
that allows an exchange of proteins between the membrane and the surrounding
fluid. Depending on the considered length scale they find several relevant
time scales for membrane height and density fluctuations. Surprisingly, this
relevance of different time scales in membrane height correlations has, to the
best of our knowledge, not been previously observed in experiments. 

In the present work we study the two interrelated effects following from a
protein-membrane interaction: the reduction of the lateral diffusion
coefficient of the protein and the modifications of the static and time
dependent height correlation functions of the membrane. In our study the
additional energy caused by the insertion of the protein arises from an
effective bending rigidity and spontaneous curvature of the protein.  The
dynamics of our system is dominated by two processes: the shape fluctuations
of the membrane and the lateral diffusion of the protein. Taking into account
the hydrodynamic interaction of a membrane with the surrounding fluid, we are
able to derive a Langevin equation for the dynamics of the membrane
shape, which becomes a function of the protein's
position. Lateral diffusion of the protein is captured by another
Langevin equation that takes into account the shape of and the interaction
with the membrane. 
These two coupled equations of motions are the
starting point for our analytical calculations and their numerical
integration make out our simulation scheme. 

Using a systematic approach we present two main results for lateral diffusion:
we argue that in equilibrium the effective lateral diffusion coefficient of a
protein that interacts with the membrane shape is universally decreased
compared to the free diffusion coefficient applicable if no interaction were
present. Beyond this general argument we, furthermore, derive an explicit
expression for the effective diffusion coefficient of a protein with
spontaneous curvature and bending rigidity by applying a path integral
approach.  To lowest order, our expression agrees with that recently derived
by Naji et al.~\cite{Naji:2009} through an estimate of the power loss of the
diffusing protein in the limit that the membrane shape minimizing the system's
energy instantaneously tracks the protein's position. Our approach reveals
that their expression resulting from a phenomenological approximation formally
corresponds to the lowest order of an expansion in $(\beta\kappa)^{-1}$ using
$\beta\equiv(k_BT)^{-1}$ with temperature $T$, Boltzmann's constant
$k_{\text{B}}$, and bending rigidity $\kappa$ of the membrane.

We compare results of our simulation scheme with this lowest order expression
and find good quantitative agreement. Several correlation functions contribute
to the effective diffusion coefficient. The quantitative analysis of these
contributions from our simulation results shows that the correlations between
the response of the membrane and the preceding stochastic force acting on the
protein effectively reduce the diffusion coefficient, while all other
contributions would cause an increase.

Concerning the altered membrane spectrum we analytically develop an
approximate expression for the height correlation function applicable for
equal bending rigidity of protein and membrane. In the limit of slow protein
diffusion compared to membrane fluctuations we find two time regimes for the
decay of height correlations: at small times the decay is dominated by
membrane dynamics, while the diffusive time scale of the protein becomes the
only relevant time scale at later times. Since we find this feature in the
simulations not only for equal bending rigidities of membrane and protein, we
suggest that the experimental analysis of the late decay of dynamical membrane
height correlations can provide a means to determine the effective diffusion
coefficient of proteins in the membrane.  To corroberate this argument we
use our model to give a rough estimate suggesting that the effect should be
visible in realistic systems. 

The paper is organised as follows: in the next section we introduce the model
of the system and develop the equations of motion both for the protein and the
membrane. In the limit of small ratios of
temperature and bending rigidity these are then used in secs.~\ref{sec:D_eff}
and~\ref{sec:height_correlations} to develop  an exact analytical expression for the
effective lateral diffusion coefficient and for the dynamical membrane height
correlations, respectively.  In sec.~\ref{sec:simulations} we briefly explain
our simulation scheme and motivate the parameters used in the simulations.  In
sec.~\ref{sec:results_diff} we show that the derived analytical
expression for the effective diffusion coefficient shows good agreement with
simulations using parameters of typical experiments. We, furthermore,
quantitatively analyse the correlation functions that contribute to the
diffusion coefficient.  The membrane height correlations are compared with our
analytical expressions in sec.~\ref{sec:results_corr}. In
sec.~\ref{sec:diff_spec} we show the
determination of the effective diffusion coefficient from the late time decay
of height correlations and discuss that this procedure should be
experimentally feasible in realistic system. We finally close with some
conclusions.
\section{Model}
\label{sec:model}
In our model we consider a single diffusing inclusion in a membrane with
bending rigidity $\kappa$ that we describe in the Monge gauge. The small
inclusion with radius $a_p$ has a spontaneous curvature $C_p$ and its
stiffness may differ from that of the membrane by a factor of $\gamma$. The
energy of the system of size $L^2$ may be expressed by
\begin{multline}
\mathcal{H}[h,\mathbf{R}]=\frac{\kappa}{2}\int_{L^2}\!d^2r\,\left\{\left(\nabla_{\mathbf{r}}^2h\right)^2+\right.\\\left.\pi
  a_p^2G(\mathbf{r}-\mathbf{R})\left[\gamma\left(\nabla_{\mathbf{r}}^2h-C_p\right)^2-\left(\nabla_{\mathbf{r}}^2h\right)^2\right]\right\}\,,
\label{eq:hamiltonian}
\end{multline}
with the height function $h(\mathbf{r})$ that quantifies the distance between the
membrane and the position $\mathbf{r}$ on a flat reference plane. The particle position
projected onto this plane is given by $\mathbf{R}$. The
function $G(\mathbf{r}-\mathbf{R})$ defines a weighting function of the
particle that must fulfil the normalization $\int_{L^2} d^2r\,G(\mathbf{r})=1$. In our
simulations we set the weighting function to a Gaussian
$G(r)=(\pi a_p^2)^{-1}\exp\left[-r^2/a_p^2\right]$ such that the transition from membrane
to particle is smooth. 
If we use the Fourier expansion
$h(\mathbf{r})=\frac{1}{L^2}\sum_\mathbf{k}h(\mathbf{k})\exp(i\mathbf{k}\cdot\mathbf{r})$
and $h(\mathbf{k})=\int_{L^2}\!d^2r\, h(\mathbf{r})\exp(-i\mathbf{k}\cdot\mathbf{r})$
the Hamiltonian becomes
\begin{multline}
 \mathcal{H}[h(\mathbf{k}),\mathbf{R}]=\frac{\kappa}{2}\left\{\frac{1}{L^2}\sum_\mathbf{k}k^4h(\mathbf{k})h(-\mathbf{k})+\right.\\
\left.(\gamma-1)\frac{\pi a_p^2}{L^4}\sum_{\mathbf{k},\mathbf{k}'}k^2k'^2G(-\mathbf{k}-\mathbf{k}')e^{i(\mathbf{k}+\mathbf{k}')\cdot\mathbf{R}} h(\mathbf{k})h(\mathbf{k}')+\right.\\
\left.2\gamma C_p\frac{\pi
  a_p^2}{L^2}\sum_\mathbf{k}k^2G(-\mathbf{k})e^{i\mathbf{k}\cdot\mathbf{R}}h(\mathbf{k})\right\}+
\gamma\pi a_p^2C_p^2\,.
\label{eq:hamiltonian_k}
\end{multline}
From this Hamiltonian it is in principle possible to numerically calculate the
equilibrium height correlations $\langle h(\mathbf{k})h(\mathbf{k}')\rangle$
applying methods used in~\cite{Merath:2006}. While these methods are
restricted to time independent equilibrium quantities, our simulation scheme,
described later in the paper, allows us to not only obtain these quantities
but also time dependent information.  For the special case that both
the protein and the membrane have the same bending rigidity, i.e.~$\gamma=1$,
the height correlations are given by
\begin{equation}
\langle
h(\mathbf{k})h(-\mathbf{k})\rangle=\frac{L^2}{k^4}\left[\frac{1}{\beta\kappa}+\rho\,\pi
    a_p^2C_p^2G(\mathbf{k})G(-\mathbf{k})\right]\,,
\label{eq:corr_eq}
\end{equation}
with the ratio of protein area to system size
\begin{equation}
\rho\equiv\pi a_p^2/L^2\ .
\label{eq:density}
\end{equation}  
While eq.~\eqref{eq:corr_eq} is derived for a single protein on the membrane
the extension to several noninteracting proteins would lead to the same result
with $\rho$ resembling the overall area density of the proteins.  Compared to
the free membrane without protein, whose height correlations are given by the
first term, the protein gives rise to an \emph{additive} term that depends on
the various parameters characterizing the particle. Following previous
studies~\cite{Leibler:1986,Divet:2002,Bivas:2003}, it is possible to define a
$k$-dependent effective bending rigidity $\kappa_{\text{eff}}(k)$ such that
the spectrum of the membrane has the form of a protein-free membrane $\langle
h(\mathbf{k})h(-\mathbf{k})\rangle_{\text{free}}=L^2/(\beta\kappa_{\text{eff}}(k)k^4)$. This
leads to
\begin{equation}
\frac{\kappa_{\text{eff}}(k)}{\kappa}=\left[1+\beta\kappa\, \rho\,\pi
    a_p^2C_p^2\,G(\mathbf{k})G(-\mathbf{k}) \right]^{-1}
\label{eq:kappa_eff}
\end{equation}
Note, that the addition of the protein in the membrane always leads to a reduction of
the effective bending rigidity of the system as has been pointed out
previously when inclusions are inserted into a
membrane~\cite{Leibler:1986}. However, the effective bending rigidity cannot
become negative, hence the membrane not unstable.

Neglecting any geometric effects caused by the projection of the protein path,
which we have previously determined to be rather small for realistic
membranes~\cite{Reister:2005}, the diffusive motion of the protein and the
thermal fluctuations of the membrane, i.e.~the dynamics of the height modes
$h(\mathbf{k},t)$, are appropriately described by the following coupled
Langevin equations:
\begin{eqnarray}
\dot{\mathbf{R}}(t)&=&-\mu_{\text{p}}\nabla_{\mathbf{R}} \mathcal{H}+\text{\boldmath $\zeta$}(t)\label{eq:Langevin1}\\
\dot{h}(\mathbf{k},t)&=&-\Lambda(k)\frac{\delta \mathcal{H}}{\delta h(\mathbf{k},t)}+\xi(\mathbf{k},t)\label{eq:Langevin3}
\end{eqnarray}
with the stochastic forces {\boldmath $\zeta$}$(t)$ and $\xi(\mathbf{k},t)$ that are related to the
mobilities $\mu_{\text{p}}\equiv D_{\text{p}}/k_BT$ of the protein~\cite{footnote}
and $\Lambda(k)$ of the membrane,
respectively, via the fluctuation-dissipation-theorems
\begin{equation}
\begin{array}{rcl}
\langle\zeta_l(t)\rangle&=&0\\
\langle\zeta_l(t)\zeta_m(t')\rangle&=&2D_{\text{p}}\,\delta_{l,m}\delta(t-t')\,,
\end{array}
\label{eq:FDT_R}
\end{equation}
and
\begin{equation}
\begin{array}{rcl}
\langle\xi(\mathbf{k},t)\rangle&=&0\\\langle\xi(\mathbf{k},t)\xi(\mathbf{k'},t')\rangle&=&2k_BT\Lambda(k)\,L^2\delta_{\mathbf{k}',-\mathbf{k}}\delta(t-t')\,.
\end{array}
\label{eq:fdt_memb}
\end{equation}
The mobility of the membrane takes into account the dynamics of the membrane
caused by the surrounding fluid. A hydrodynamical derivation involving the
Oseen tensor leads to a mobility of $\Lambda(k)=(4\eta k)^{-1}$ for the
undulations $k\neq 0$, with the viscosity $\eta$ of the surrounding
fluid~\cite{Seifert:1997}. For $k=0$, i.e.~the center of mass movement of the
membrane, we set $\Lambda(k=0)=0$ since it does not influence the properties
of interest in our study.
\section{Analytical approach}
\subsection{Diffusion coefficient $D_{\text{eff}}$ }
\label{sec:D_eff}
We derive an analytical expression for the effective diffusion coefficient
$D_{\text{eff}}$ by exploiting that for biomembranes typically
$(\beta\kappa)^{-1}\ll 1$.

We first determine the minimum of the
energy~\eqref{eq:hamiltonian_k}. 
The condition  $\left.\frac{\partial\mathcal{H}}{\partial
  h(\mathbf{k})}\right|_{\hat{h}_{\mathbf{k}}}=0$ leads to the equation
\begin{multline}
0=k^2\hat{h}_{-\mathbf{k}}+\\(\gamma-1)\rho \sum_{\mathbf{k}'}k'^2G(\mathbf{k}+\mathbf{k}')\exp[i(\mathbf{k} +
\mathbf{k}') \cdot\mathbf{R}]\hat{h}_{\mathbf{k}'}\\+\gamma C_p \pi a_p^2 G(\mathbf{k})
\exp(i\mathbf{k} \cdot \mathbf{R})\ ,
\end{multline}
for the height modes $\hat{h}_{\mathbf{k}}$ that minimize the energy.
Using the ansatz $\hat{h}_\mathbf{k} = \frac{B_\mathbf{k}}{k^2}
\exp(-i\mathbf{k}\!\cdot\!\mathbf{R})$  the energy is
minimal for
\begin{equation}
B_{\mathbf{k}}=-\gamma C_p\pi a_p^2
\sum_{\mathbf{q}}M_{\mathbf{k,q}}^{-1}G(\mathbf{q})\ ,
\label{eq:B_k}
\end{equation}
with the matrix
\begin{equation}
M_{\mathbf{k,q}}\equiv\delta_{\mathbf{k,q}}+(\gamma-1)\rho\, G(\mathbf{q}+\mathbf{k})\ .
\end{equation}
Inserting this result into the 
Hamiltonian shows that the energy minimum does
not depend on the particle position as expected from the isotropy of the particle position. 

The first question we will address is whether the effective diffusion constant
is larger or smaller than the free diffusion coefficient $D_{\text{p}}$
applicable without coupling, i.e.~$\gamma=0$.  The degrees of freedom in the
Hamiltonian~\eqref{eq:hamiltonian_k} are given by the membrane modes
$h(\mathbf{k})$ and the position $\mathbf{R}$ of the protein. Due to the
appearance of the membrane modes up to quadratic order in the Hamiltonian, the
thermal averages $\langle |h(\mathbf{k})|^2\rangle$ remain bounded.  The
position of the protein, however, is not bounded, such that diffusive motion
is possible. Effectively, the protein moves in a time-dependent periodic
potential given by the height modes $h(\mathbf{k},t)$.  Diffusion in periodic
potentials has been previously considered in a large number of studies. If the
particle is only subject to the potential and no other external force it is
easily shown, that the effective diffusion coefficient of the particle is
always smaller than or equal to the free diffusion
coefficient~\cite{Lifson:1962,Reimann:2002}. Thus we conclude that the
effective diffusion coefficient $D_{\text{eff}}$ of a protein whose
interaction with the membrane depends on the shape obeys
\begin{equation}
D_{\text{eff}}\leqslant D_{\text{p}}
\label{eq:statement}
\end{equation}
in all situations without external driving forces. 

While we know that the diffusion coefficient is in general reduced due to
the membrane we will derive an explicit analytical expression to quantify the
effect for our model, the validity of which we will discuss and analyze through
simulations.  A quick glance at the Langevin equations~\eqref{eq:Langevin1}
and \eqref{eq:Langevin3}
shows that they are highly nonlinear such that the exact solution is not
straightforward. In order to develop an expression for $D_{\text{eff}}$ we
must, therefore, apply certain approximations.

In a path integral description ~\cite{Risken:1996} the probability
distribution $\mathcal{P}$ of the paths $\mathbf{R}(t)$ of the diffusing
protein and of the height modes $h(\mathbf{k},t)$ follow from the weight of
noise fluctuations and are given by the functional
\begin{equation}
\mathcal{P}[\mathbf{R}(t),h(\mathbf{k},t)]\sim\exp\left[-\frac{1}{2k_BT}\int_0^t\!d\tau
  L(\mathbf{R}(\tau),h(\mathbf{k},\tau))\right],
\label{eq:path}
\end{equation}
with the function
\begin{multline}
  L(\mathbf{R},h(\mathbf{k}))=\frac{1}{2\mu_0}\left(\dot{\mathbf{R}}+\mu_0\nabla_{\mathbf{R}}\mathcal{H}
  \right)^2\\+\sum_{\mathbf{k}}\frac{1}{2\Lambda(k)}\left|\dot{h}(\mathbf{k})+\Lambda(k)\frac{\partial\mathcal{H}}{\partial
      h(\mathbf{k})}\right|^2\ .
\end{multline}
If we introduce the deviation 
\begin{equation}
y_\mathbf{k}\equiv h(\mathbf{k})-\hat{h}_\mathbf{k}(\mathbf{R}(t))\,,
\end{equation}
 of the membrane shape from the instantaneous membrane shape that minimizes
 the energy of the system the Hamiltonian may approximately be written in the form
\begin{equation} 
\mathcal{H}=
\mathcal{H}_0+\frac{1}{2}\sum_{\mathbf{k},\mathbf{k}'}y_\mathbf{k}\left.\frac{\delta^2\mathcal{H}}{\delta
  h(\mathbf{k})\delta h(\mathbf{k}')}\right|_{\hat{h}_\mathbf{k}(\mathbf{R})}y_{\mathbf{k}'}\ ,
\end{equation}
where $\mathcal{H}_0\equiv\mathcal{H}[\hat{h}(t)]$ is the energy
minimum. Since the second functional derivative of the energy with respect to
the height is proportional to the bending rigidity $\kappa$, it is convenient
to define
\begin{equation}
\beta\kappa\, V_{\mathbf{k},\mathbf{k}'}(\mathbf{R})\equiv\left.\frac{\delta^2\mathcal{H}}{\delta
  h(\mathbf{k})\delta h(\mathbf{k}')}\right|_{\hat{h}_\mathbf{k}(\mathbf{R})}\ .
\end{equation}
Up to total derivatives leading only to boundary terms we can then rewrite
the function $L$ as a function of $\mathbf{R}$ and $y_\mathbf{k}$
\begin{multline}
L(\mathbf{R},y(\mathbf{k}))=\frac{1}{2\mu_0}\dot{\mathbf{R}}^2+\\ 
\sum_{\mathbf{k}}\frac{1}{2\Lambda(k)}
\left[\left|(\dot{\mathbf{R}}\cdot\nabla_{\mathbf{R}})  \hat{h}_\mathbf{k}\right|^2
  +2\dot{y}^*_\mathbf{k}(\dot{\mathbf{R}}\cdot\nabla_{\mathbf{R}}) \hat{h}_\mathbf{k}+
  \left|\dot{y}_\mathbf{k}\right|^2\right]\\ 
+\frac{1}{2}\left[\mu_0(\beta\kappa)^2\bigg(\sum_{\mathbf{k},\mathbf{k}'}
  (y_\mathbf{k}\nabla_{\mathbf{R}}V_{\mathbf{k},\mathbf{k}'}y_{\mathbf{k}'}
  -y_\mathbf{k}V_{\mathbf{k},\mathbf{k}'}\nabla_{\mathbf{R}}\hat{h}_{\mathbf{k}'}
  )\bigg)^2\right.\\
  +\left.\frac{1}{2}(\beta\kappa)^2\sum_{\mathbf{k}}\Lambda(\mathbf{k})\bigg(\sum_{\mathbf{k}'}V_{\mathbf{k},\mathbf{k}'}y_{\mathbf{k}'}\bigg)^2\right]\,.
\label{eq:onsager}
\end{multline}
To determine the effective mobility of the particle it
would be  necessary to integrate out the deviation $y$ in the probability
distribution $\mathcal{P}$. Since this cannot be done explicitly, we employ a
saddle point approximation, i.e.~we replace all the possible paths
$y_{\mathbf{k}}(t)$ with the path $\tilde{y}_{\mathbf{k}}(t)$ that minimizes
the function $L$ and hence contributes to the probability
distribution~\eqref{eq:path} the most. The path $\tilde{y}_{\mathbf{k}}(t)$
follows from the Euler-Lagrange-equations 
\begin{equation}
\frac{d}{dt}\frac{\partial
  L}{\partial \dot{\tilde{y}}_{\mathbf{k}}}-\frac{\partial L}{\partial
  \tilde{y}_{\mathbf{k}}}=0\,.
\end{equation}
 As before,  we assume that only small variations $\tilde{y}_{\mathbf{k}}$
in the height are relevant allowing us to linearize the resulting differential
equations
\begin{multline}
\frac{1}{\Lambda(\mathbf{k})}\ddot{\tilde{y}}_\mathbf{k}(t)-(\beta\kappa)^2
\sum_{\mathbf{k}''}\Lambda(\mathbf{k}'')
\bigg(\sum_{\mathbf{k}'}
V_{\mathbf{k}'',\mathbf{k}'}\tilde{y}_{\mathbf{k'}}\bigg)V_{\mathbf{k}'',\mathbf{k}}\\
-\mu_0(\beta\kappa)^2\bigg(\sum_{\mathbf{k}'',\mathbf{k}'}\nabla_{\mathbf{R}}\hat{h}_\mathbf{k'}V_{\mathbf{k}'',\mathbf{k}'}\tilde{y}_{\mathbf{k}''}\bigg)
\sum_{\mathbf{k}'}V_{\mathbf{k},\mathbf{k}'}\nabla_{\mathbf{R}}\hat{h}_\mathbf{k'}\\
=-\frac{1}{\Lambda(\mathbf{k})}\frac{d^2\hat{h}_\mathbf{k}(\mathbf{R}(t))}{dt^2}\ ,
\end{multline}
or, in a  simplified notation,
\begin{equation}
\ddot{\tilde{y}}_\mathbf{k}(t)-(\beta\kappa)^2\sum_{\mathbf{k},\mathbf{k}'}A_{\mathbf{k},\mathbf{k}'}\tilde{y}_{\mathbf{k}'}=-\frac{d^2\hat{h}_\mathbf{k}(\mathbf{R}(t))}{dt^2}\ ,
\label{eq:de_y}
\end{equation}
using the positive definite matrix $A_{\mathbf{k},\mathbf{k}'}$. The
homogeneous solution of this equation is a simple relaxation on time scales
proportional to $(\beta\kappa)^{-1}$ and plays, therefore, a minor role for large values
of $\beta\kappa$. We are now interested in the significance of the bending rigidity
$\kappa$ on the inhomogeneous solution.  To this end we drop all dependencies
of $\mathbf{k}$ in eq.\eqref{eq:de_y} as though the system only had a single wave
mode without altering the order of $\beta\kappa$.
The inhomogeneous solution is then given by
\begin{equation}
\tilde{y}_{\text{inh}}(t)=\frac{1}{2\beta\kappa\sqrt{A}}\int_0^t\!d\tau\left(e^{\beta\kappa\sqrt{A}(t-\tau)}-e^{\beta\kappa\sqrt{A}(\tau-t)}\right)\frac{d^2\hat{h}}{d\tau^2}.
\end{equation}
For large $\beta\kappa$ and slow protein diffusion the membrane shape minimizing
the energy $\hat{h}(\mathbf{R}(t))$ only weakly changes on the relaxation time
scale. Hence, membrane shape deviations $\tilde{y}_{\mathbf{k}}$ are of the order $\mathcal{O}((\beta\kappa)^{-2})$.
Inserting this result into eq.~\eqref{eq:onsager} and keeping only leading
orders of $(\beta\kappa)^{-1}$ we arrive at
\begin{multline}
L(\mathbf{R},\tilde{y}(\mathbf{k}))=\\\frac{1}{2\mu_0}\dot{\mathbf{R}}^2+
\sum_{\mathbf{k}}\frac{1}{2\Lambda(k)}\left|(\dot{\mathbf{R}}\cdot\nabla_{\mathbf{R}})
\hat{h}_\mathbf{k}\right|^2+\mathcal{O}((\beta\kappa)^{-2})\ .
\end{multline}
It is now possible to identify an effective diffusion coefficient
$D_{\text{eff}}$ for the diffusing protein from the prefactor of the
$\dot{\mathbf{R}}^2$ term
\begin{eqnarray}
\frac{D_0}{D_{\text{eff}}}&=&1+\sum_\mathbf{k}\frac{\mu_0}{\Lambda(k)}\left|\nabla_{\mathbf{R}} \hat{h}_\mathbf{k}\right|^2\nonumber\\
&=&
1+\sum_\mathbf{k}\frac{\mu_0}{\Lambda(k)}\frac{B_{\mathbf{k}}B_{-\mathbf{k}}}{k^2}\,,
\label{eq:result}
\end{eqnarray}
with $B_{\mathbf{k}}$ from eq.~\eqref{eq:B_k} and using isotropy in the $x$-
and $y$-direction of the system.  This systematic derivation of an analytical expression
for the diffusion of a protein that interacts with the membrane shape
constitutes our first main result. The analysis of experiments with both model
and biological membranes, shows, that $(\beta\kappa)^{-2}$ is typically
smaller than $0.01$, hence, sufficiently small to expect a wide applicability
of this expression. The first line of this expression agrees with the result
of Naji et al.~\cite{Naji:2009}. In their derivation they apply an adiabatic
approximation assuming the membrane shape to instantaneously follow the path
of the protein. The effective diffusion coefficient is then derived from an
estimation of the power loss of the diffusing particle. Our derivation
identifies their approximate result as the lowest order of an expansion in
$(\beta\kappa)^{-1}$.

After the general solution for the effective diffusion coefficient we will now
turn to the special case of a protein with a  weighting function expressed
through Dirac's delta-function
$G(\mathbf{R}-\mathbf{r})=\delta(\mathbf{R}-\mathbf{r})$. 
In this case the Fourier-transform of $G$ is independent of the wave-vector
$\mathbf{k}$, such that $G(\mathbf{k})=1$. This leads to the height-mode
\begin{equation}
\hat{h}_{\mathbf{k}}(\mathbf{R})=-\frac{1}{k^2}\frac{\gamma C_p\pi
  a_p^2}{1+(\gamma-1)\rho\sum_{\mathbf{k}'}1}\exp(-i\mathbf{k}\cdot\mathbf{R})\,,
\end{equation}
minimizing the free energy $\mathcal{H}$. The resulting energy minimum is
\begin{equation}
\mathcal{H}[\hat{h}]=
\frac{\kappa}{2}\gamma\pi a_p^2
C_p^2\frac{1}{1+\gamma\rho\left(\sum_{\mathbf{k}}1\right)/\left(1-\rho\sum_{\mathbf{k}}1\right)}\;.
\end{equation}
Following the above procedure but now inserting the special choice for the
weighting function leads to the effective diffusion coefficient
\begin{equation}
\frac{D_{0}}{D_{\text{eff}}}=1+\sum_{\mathbf{k}}\frac{\mu_0}{\Lambda(k)}\frac{1}{k^2}\frac{\gamma^2C_p^2\pi^2a_p^4}{\left(1+(\gamma-1)\rho\left(\sum_{\mathbf{k}'}1\right)\right)^2}\ . 
\label{eq:diff_delta}
\end{equation}
In a next step we will evaluate the sums over all possible wave-vectors
$\mathbf{k}$. On the one hand the smallest value of the $x$ or $y$ component ist defined by
the system size, $k_{x,\text{min}}=2\pi/L$, and is thus approximately zero for
large systems. On the other hand the largest value $k_{x,\text{max}}=2\pi/a_c$ is
limited by a microscopic cutoff length $a_c$ that corresponds to the size of
the lipids. Thus the expression $\rho\sum_{\mathbf{k}}1$ in the denominator of
eq.~\eqref{eq:diff_delta} leads to $\pi a_p^2/a_c^2$ using the definition of
$\rho$ given in eq.~\eqref{eq:density}. The evaluation of the other sum over
$\mathbf{k}$ depends on the specific form of the membrane mobility. Using
$\Lambda(k)=(4\eta k)^{-1}$ makes the evaluation of $\sum_{\mathbf{k}}k^{-1}$
necessary, 
such that the expression for the ratio of the diffusion coefficients becomes
\begin{equation}
\frac{D_0}{D_{\text{eff}}}=1+\frac{\mu_04\eta\gamma^2C_p^2\pi
  a_p^4}{\left(1+(\gamma -1)\frac{a_p^2}{a_c^2}\right)^2}
\ln\left(\frac{\sqrt{2}+1}{\sqrt{2}-1}\right)\frac{L^2}{a_c}\,.
\end{equation}
Typically the stiffness of a diffusing protein will be significantly larger
than that of the membrane. In the limit $\gamma\gg 1$ the resulting effective
diffusion coefficient is given by the relation
\begin{equation}
\frac{D_0}{D_{\text{eff}}}=1+ \mu_04\eta C_p^2\pi
\ln\left(\frac{\sqrt{2}+1}{\sqrt{2}-1}\right)L^2a_c^3\,.
\end{equation}

Equation \eqref{eq:diff_delta} shows that a non-zero spontaneous curvature $C_p$ is
crucial in order to have an influence on the diffusion coefficient of the protein. Within
our model a mere difference in the bending rigidity of the membrane and the
protein, i.e.~$\gamma\neq1$, does not lead to an altered diffusion coefficient.

Before testing our expression by comparing it with simulations, we will discuss
two limiting cases for eq.~\eqref{eq:diff_delta}. Without altering the general
conclusions we will give this discussion only for a single $k$-mode. As was
expected, our expression reveals an effective diffusion coefficient that
always has an upper bound of $D_0$. For very small ratios $\mu_0/\Lambda$
($\Lambda\equiv\Lambda(k)$), i.e.~if the membrane is much more mobile than the
protein, the reduction of the effective diffusion coefficient is linear in
this ratio $D_{\text{eff}}/D_0\approx 1-|B_\mathbf{k}|^2\,\mu_0/\Lambda$. The
free mobility of the protein dominates its effective movement. The influence
of the membrane is weak since it can adjust quickly to the position of the
protein. In this situation our approximation that the system's energy is
always close the minimum is fulfilled and our expression will serve as a very
good estimate.  For the limit in which the membrane moves much slower than the
particle, which corresponds to the scenario of diffusion in an (almost) fixed
periodic potential, our expression predicts the asymptotic behavior
$D_{\text{eff}}/D_0 \approx |B_\mathbf{k}|^{-2}\Lambda/\mu_0$. Thus the
movement of the protein is mainly dominated by the membrane mobility such that
it is strongly slowed down. However, if the diffusing particle effectively
sees a fixed energy landscape the stochastic motion enables the protein to hop
from one energy minimum to another, such that our previous approximation that
the protein always stays very close to the position of the instantaneous
energy minimum may no longer be valid.
\subsection{Temporal decay of membrane height correlations}
\label{sec:height_correlations}
Using the equations of motion~\eqref{eq:Langevin1} and~\eqref{eq:Langevin3},
a calculation of the full height  correlation function
$\langle h(\mathbf{k},t)h(\mathbf{k}',t')\rangle$  is not feasible
analytically. We rather determine this quantity from our simulation
scheme. Nevertheless, in order to gain an understanding of the possible
contributions to the correlation function, it is instructive to consider the
special case of equal bending rigidities of the particle and the membrane,
$\gamma=1$.  The general solution of the
Langevin equation~\eqref{eq:Langevin3} is then given by
\begin{equation}
  \begin{split}
  h(\mathbf{k},t)=&h(\mathbf{k},0)e^{-t/\tau_M(k)}
  +e^{-t/\tau_{M}(k)}\\&\int_0^t\!dt'
  e^{t'/\tau_M(k)}\left[\xi(\mathbf{k},t')-\frac{C_p\pi a_p^2G(\mathbf{k})}{k^2\tau_M(k)}e^{-i\mathbf{k}\cdot\mathbf{R}(t')}\right]
\end{split}
\end{equation}
with the $k$-dependent membrane time scale 
\begin{equation}
\tau_M(k)\equiv 4\eta/(\kappa k^3)\,.
\end{equation}
 Using the fluctuation-dissipation theorem ~\eqref{eq:fdt_memb} and
assuming that the particle diffuses on timescales much larger than $\tau_M(k)$
the height correlation function is given by
\begin{multline}
\langle
h(\mathbf{k},t)h(-\mathbf{k},t')\rangle=\frac{L^2}{\beta\kappa k^4}\left[e^{-|t-t'|/\tau_M(k)}+\right.\\\left.
\beta\kappa\,\rho\,\pi a_p^2C_p^2G(\mathbf{k})G(-\mathbf{k})e^{-|t-t'|/\tau_D(k)}\right]
\label{eq:corr}
\end{multline}
with the diffusive time scale 
\begin{equation}
\tau_D(k)\equiv(D_{\text{eff}}k^2)^{-1}\,.
\end{equation}
In order to arrive at eq.~\eqref{eq:corr} we use
$\langle\exp[-i\mathbf{k}\cdot(\mathbf{R}(t)-\mathbf{R}(t'))]\rangle=e^{-|t-t'|/\tau_D(k)}$
which follows for diffusive motion of the protein with an effective
diffusion coefficient $D_{\text{eff}}$.

Equation~\eqref{eq:corr} shows that the dynamics of the height correlation
function of the membrane is determined by the two timescales present in the
system: the membrane time scale $\tau_M(k)$ and the diffusive time scale
$\tau_D(k)$. Our calculations assumed $\tau_M(k)<\tau_D(k)$, thus, the decay
of height correlations for small times $t-t'$ will be dominated by the
membrane dynamics while for large times the diffusion of the particle takes
over. While our calculation is strictly valid only in the case of $\gamma=1$
the qualitative behaviour persists also for the case $\gamma\neq1$ as will be
shown when we present simulation results. Note, that the naive usage of an
effective binding rigidity $\kappa_{\text{eff}}(k)$ would lead to a single
timescale for each mode $k$. Since the properties of the system are clearly
dominated by two timescales the concept of $\kappa_{\text{eff}}(k)$ is only
applicable for properties that are not time dependent.
\section{Simulations}
\subsection{Scheme} 
\label{sec:simulations}
Our simulation scheme comprises the numerical integration of the two
coupled Langevin equations \eqref{eq:Langevin1} and
\eqref{eq:Langevin3}. However, the equation of motion given for the protein in
eq.~\eqref{eq:Langevin1} neglects that the particle actually diffuses along
the membrane, in other words a curved surface. The shape of the membrane
influences the Langevin equation, the exact form of which is given in
refs.~\cite{Leitenberger:2007a,Reister:2007} and used in our simulations. Thus
the free diffusion coefficient $D_0$ used in the simulations is slightly
larger than the value of $D_\text{p}$ in eq.~\eqref{eq:Langevin1}~\cite{footnote}.

The membrane is mapped on a square $N\times N$-lattice such that the length of
the system is $L=N\,\ell$ with the lattice spacing $\ell$.  The membrane shape
is evolved in time by a time discrete version of eq.~\eqref{eq:Langevin3} in
Fourier space. This part of the scheme is an extension of the Fourier Space
Brownian Dynamics simulation method introduced by Lin and
Brown~\cite{Lin:2004,Lin:2005,Brown:2008}.

After every update of the membrane shape the position of the
particle is altered by using a discrete version of
eq.~\eqref{eq:Langevin1}. However, the particle's position is not evolved on
the lattice. The membrane height at the particle position that enters in the
equation of motion is determined through linear extrapolation from the height
at the four nearest neighbour lattice sites. The shape of the membrane in real
space $h(\mathbf{r},t)$ is determined by use of fast Fourier transforms
implementing the FFTW library~\cite{FFTW05}.  For a more detailed account of
the simulation scheme we refer the reader to
refs.~\cite{Leitenberger:2007a,Reister:2007}.

All simulation results presented in this paper were performed on a
$64\times64$-lattice with a lattice spacing of $\ell=10$nm. The radius of the
protein is set to $a_p=2\ell$. The fluid surrounding the membrane is water
with a viscosity of $\eta=10^{-3}$kg/(m$\,$s) or
$\eta=2.47\times10^{-7}$s/($\beta\ell^2$) in the units of our model at $T=300
K$. The discrete integrations of both membrane shape and particle position are
performed with a timestep of $\Delta t=10^{-9}$s that is significantly smaller
than the smallest time scale $\tau_{M,\text{min}}$ in the system. If not
stated otherwise the diffusion coefficient of the protein is set to
$D_0=5\times10^{-8}$cm$^{-2}$/s$=5\times10^{4}\ell^2$/s. This ensures that the membrane time scale
is always smaller than the diffusive timescale $\tau_{M}(k)<\tau_D(k)$ as is
the case in real biological systems. Simulation runs were performed with
$8\times10^6$ integration steps resulting in trajectories that last for $8$ms,
which is approximately five times the longest membrane time scale
$\tau_{M,\text{max}}$. The graphs presented in the following are the results
of averaging over a minimum of 500 independent trajectories, where the first
$10^6$ timesteps were not taken into account in order to ensure equilibration
of the membrane configuration and the particle position relative to the
membrane shape.
\subsection{The effective diffusion coefficient}
\label{sec:results_diff}
\begin{figure}
\includegraphics[width=0.8\linewidth]{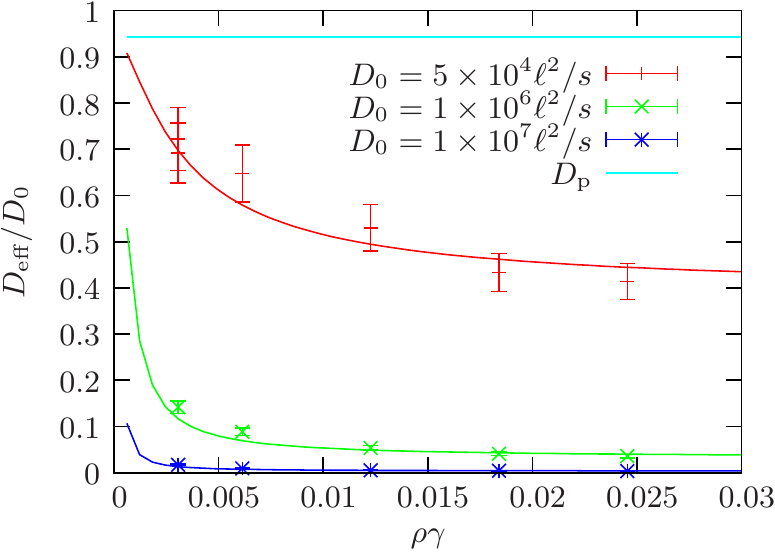}
\caption{\label{fig:stefan} Ratio $D_{\text{eff}}/D_0$ as a function of the
  coupling coefficient $\rho\gamma$ for the given free diffusion coefficients
  $D_0$ in units of $\ell^2$/s. The bending
  rigidity of the membrane is $\beta\kappa=5$ and the spontaneous curvature of
  the protein $C_p\ell=1$. 
    }
\end{figure}

To test our explicit expression~\eqref{eq:result} we have performed elaborate
simulations using the scheme described in the previous section. In
fig.~\ref{fig:stefan}, we present the resulting $D_{\text{eff}}/D_0$ as a
function of $\rho \gamma$ with the ratio of protein area to system size
$\rho$, eq.~\eqref{eq:density},
and the ratio of the protein to membrane bending rigidity $\gamma$ for three
different protein mobilities.
The detailed parameters of the simulations are given in the figure
caption. 
The comparison of the simulation results and the corresponding analytical
expression shows very good agreement for all the chosen parameters.  Since the
simulations were all performed with $(\beta\kappa)^{-2}=0.04$ we are well
within the limits, where we expect our analytical result to hold. These
simulation parameters were chosen because they represent realistic parameters
for biological systems. We thus conclude that our approach to determining the
reduction of the effective diffusion coefficient will be of use in
experimental studies.

\begin{figure}
  \centering
  \includegraphics[width=0.8\linewidth]{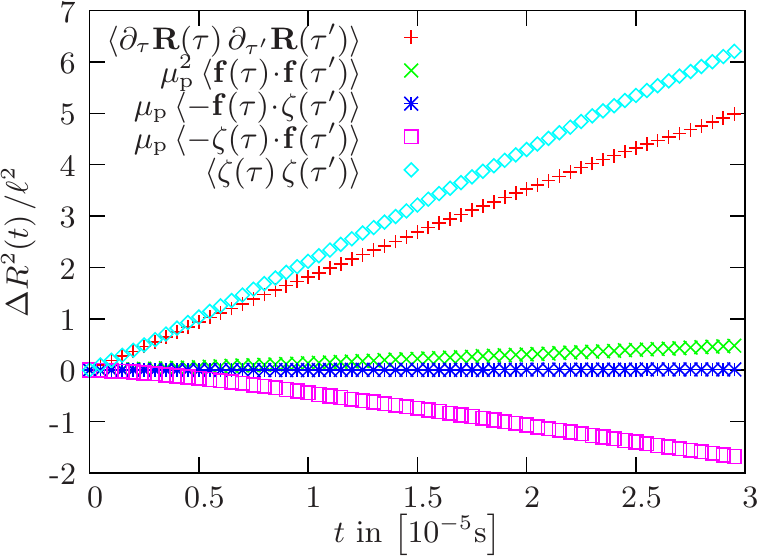}
  \caption{Force correlations integrated twice in time to show the respective
    contribution to the mean squared displacement  of the protein for
    $\gamma=1$ and $C_p\ell=1$. The results apply for $\tau\leqslant\tau'$.
    \label{fig:MSD_forcecorr}}
\end{figure}
In the following we will quantitatively analyse the contributions of the
correlation functions entering the effective diffusion coefficient. The mean
squared displacement $\langle\Delta\mathbf{R}^2(t)\rangle$ for the diffusing
protein is formally given by integrating eq.~\eqref{eq:Langevin1} twice in time
\begin{multline}
\langle\Delta\mathbf{R}^2(t)\rangle=
\int_0^td\tau\int_0^td\tau'\left\langle\partial_{\tau}\mathbf{R}(\tau)\cdot\partial_{\tau'}\mathbf{R}(\tau')\right\rangle\\
=\int_0^t d\tau\int_0^td\tau'\left[\mu_{\text{p}}^2\langle\mathbf{f}(\tau)\cdot\mathbf{f}(\tau')\rangle+\langle\text{\boldmath$\zeta$}(\tau)\cdot\text{\boldmath$\zeta$}(\tau')\rangle\right.\\\left.
+\mu_{\text{p}}\langle\mathbf{f}(\tau)\cdot\text{\boldmath$\zeta$}(\tau')\rangle+\mu_{\text{p}}\langle\text{\boldmath$\zeta$}(\tau)\cdot\mathbf{f}(\tau')\rangle\right]\,,
\end{multline}
with the conservative force
$\mathbf{f}(t)\equiv-\nabla_{\mathbf{R}}\mathcal{H}[h,\mathbf{R}]$ that the
membrane exerts on the protein. Thus the mean squared displacement has several
additive contributions. Since the effective diffusion coefficient follows from
the slope of the mean square displacement as a function of time via
$\langle\Delta\mathbf{R}^2\rangle\equiv4D_{\text{eff}}^{\text{MSD}}t$, also
the diffusion coefficient has various additive contributions.  In
fig.~\ref{fig:MSD_forcecorr}, we display the simulation results for the various
parts of the mean squared displacement for a chosen set of parameters. The
correlations of the stochastic force acting on the particle lead to a
$4D_{\text{p}}t$ behavior as is expected from the fluctuation dissipation
theorem of eq.~\eqref{eq:FDT_R}. As we have argued before, the slope of the
particle's mean squared displacement as a function of time is smaller than
$4D_{\text{p}}$, hence, one of the additive terms must be negative. However, the
force correlations $\langle\mathbf{f}(\tau)\cdot\mathbf{f}(\tau')\rangle$
obviously also lead to an additive contribution, which we find to be quite
small for the parameters of our simulations. Due to causality correlations
$\langle\mathbf{f}(\tau)\cdot\text{\boldmath$\zeta$}(\tau')\rangle$ with
$\tau\leqslant\tau'$ must be zero, such that the remaining correlations
$\langle\text{\boldmath$\zeta$}(\tau')\cdot\mathbf{f}(\tau)\rangle$ are the
cause of the reduction of the diffusion coefficient, as we clearly see from
the simulation results. This contribution expresses the reaction of the
membrane to the random force acting on the protein: If the random force moves
the particle during a small discrete timestep, the interaction of the protein
with the membrane will slightly change the shape of the membrane during the
next timestep such that the system comes closer to the energy minimum. However,
it cannot be reached during such a short time, leading to the
membrane ``pulling back'' the protein to its initial position before the
random movement. This explains the sign of the corresponding correlation
function. An important aspect here is that the membrane shape reacts to the
movement of the protein. If the membrane shape evolves independently from the
particle position these correlations do not exist leading to an increase in
the effective diffusion coefficient~\cite{Reister:2005,Leitenberger:2007a}.

\subsection{Membrane height correlations}
\label{sec:results_corr}
In the following we will elucidate that our simulation results for equal
bending rigidity of membrane and protein, $\gamma=1$, agree very well with the
analytical expressions given in eqs.~\eqref{eq:corr_eq} and
\eqref{eq:corr}. Furthermore, we will show that the qualitative features of
these equations are also observed in the more general case $\gamma\neq1$.

\begin{figure}
  \centering
  \includegraphics[width=0.8\linewidth]{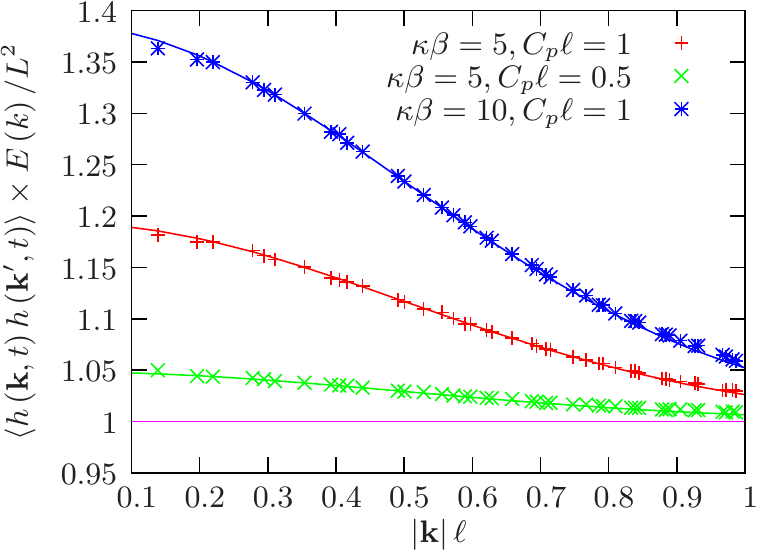}
   \caption{Mean squared height of a membrane with protein normalised by  the
    height correlations of an unperturbed membrane as a function of $k$ for equal
    bending rigidity of the protein and the membrane, $\gamma=1$. The values
    for the membrane bending rigidity $\beta\kappa$ and the spontaneous
    curvature $C_p$ are displayed in the legend. We
    define $E(k)\equiv(\beta\kappa k^4)^{-1}$.
    \label{fig:average_hh-k4-Cp}}
\end{figure}
In fig.~\ref{fig:average_hh-k4-Cp}, we present the height correlation spectrum
$\langle h(\mathbf{k})h(-\mathbf{k})\rangle$ as a function of $k^4$ for
$\gamma=1$ and different membrane bending rigidities $\kappa$ and spontaneous
curvatures $C_p$ of the protein.  In order to focus on the influence of the
protein we have normalised $\langle h(\mathbf{k})h(-\mathbf{k})\rangle$ by the
spectrum of a bare, protein-free membrane. While the symbols represent results
from the simulations the solid lines follow from eq.~\eqref{eq:corr_eq} using
the Gaussian weighting function $G(\mathbf{r}-\mathbf{R})$ given in
sec.~\ref{sec:model}. The influence of the protein is most pronounced for small wave vectors
$k$ or large length scales and decreases with increasing $k$ to the value of
the bare membrane without protein. Membrane fluctuations on length scales significantly
smaller than the inclusion's size are not influenced by the interaction of the
protein with the membrane. Comparing simulation results with the
analytical expression~\eqref{eq:corr_eq} we find that the agreement is very
good as was of course expected.
\begin{figure}
  \centering
  \includegraphics[width=0.8\linewidth]{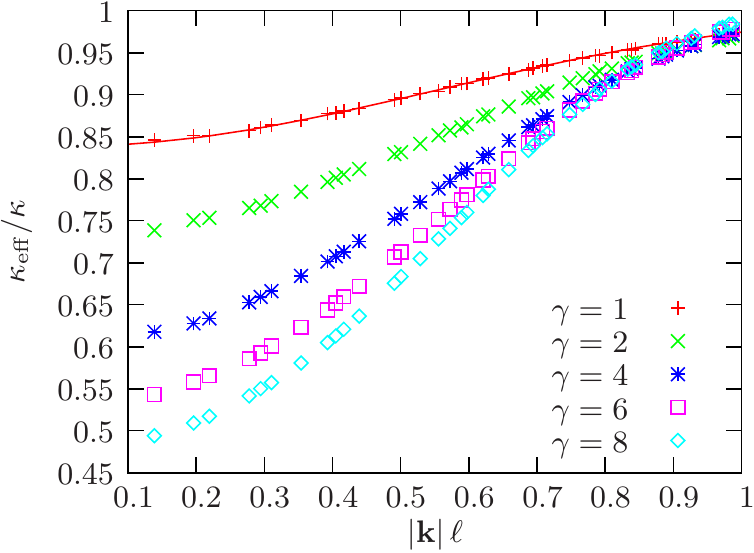}
   \caption{Effective bending rigidity $\kappa_{\text{eff}}(k)$ as a function
    of $k$ for the given values of $\gamma$. The bending rigidity of the
    membrane is $\beta\kappa=5$ and spontaneous curvature of the protein $C_p\ell=1$.
    \label{fig:kappa_eff}}
\end{figure}
In fig.~\ref{fig:kappa_eff}, we plot the effective bending rigidity
$\kappa_{\text{eff}}(k)$ as a function of $k$ as determined from the height
correlations that result from simulations with a constant spontaneous
curvature $C_p$ and membrane bending rigidity $\kappa$, but different bending
rigidity ratios $\gamma$. For $\gamma=1$ the simulation result agrees very
well with eq.~\ref{eq:kappa_eff}. With increasing the stiffness of the
particle we find that the qualitative behaviour remains similar, however,
$\kappa_{\text{eff}}$ is even more reduced. Effectively, an increase in the
bending rigidity of the protein leads to a softening of the system. Our
results indicate that  $\kappa_{\text{eff}}(k)$ saturates with increasing
$\gamma$. To corroborate this assumption simulations with even higher $\gamma$
would need to be performed, but have turned out to be very demanding
computationally.
\begin{figure}
  \centering
  \includegraphics[width=0.8\linewidth]{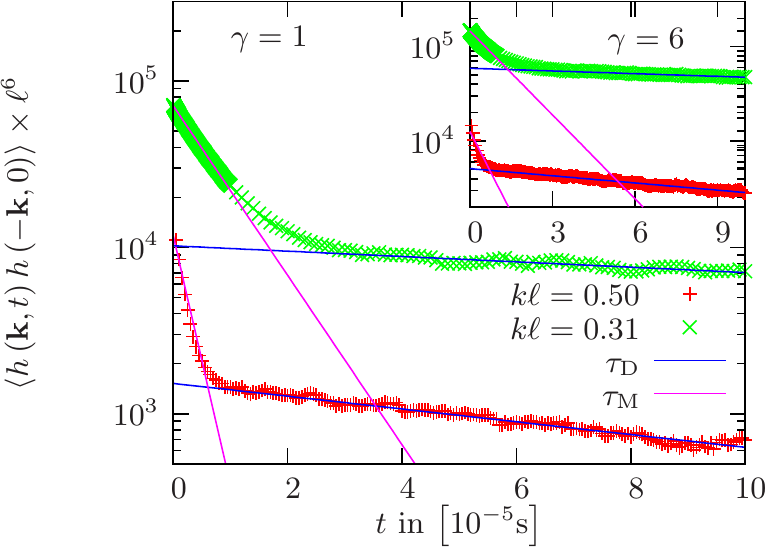}
  \caption{\label{fig:two_time_regimes} Height correlation
    functions $\langle h(\mathbf{k},t)h(-\mathbf{k},0)\rangle$ for two
    arbitrarily chosen $k$-values as a function of time $t$. The slopes of the
    solid lines determine the two time regimes dominated by
    $\tau_{\text{M}}(k)$ for small $t$ and $\tau_{\text{D}}(k)$ for large $t$.}
\end{figure}

We now turn to the temporal decay of height correlations. In
fig.~\ref{fig:two_time_regimes}, we display $\langle h(\mathbf{k},t)
h(-\mathbf{k},0)\rangle$ as a function of time for two arbitrary wave numbers
$k$. The main plot considers $\gamma=1$, the inset
$\gamma=6$. Equation~\eqref{eq:corr} suggests that the two relevant time
scales in the system, that are well separated in our calculations, become
observable: for small times the decay is dominated by the membrane time scale
$\tau_M(k)$, while for larger times the decay is predominantly influenced by
the movement of the protein and hence the corresponding time scale is
$\tau_D(k)$. Regarding the simulation results we see indeed that the behaviour
of the correlations is dominated by a fast decay at small times and a slower
decrease for large times. While eq.~\eqref{eq:corr} is an approximate result
only for $\gamma=1$ we find that this feature of two dominating time scales is
qualitatively also observed for $\gamma\neq 1$.  The quantitative fit of the
small time behaviour with an exponential decay with the characteristic time
$\tau_M(k)$ shows very good agreement for both considered values of
$\gamma$. For $\gamma\neq 1$ the Hamiltonian of the
system~\eqref{eq:hamiltonian_k} causes an additional contribution to the inverse
characteristic time that depends on $\gamma$. However, for the parameters of
our simulations this contribution is obviously negligible.

At large times the decay is expressed through the time scale $\tau_D(k)$ that
is a function of the effective diffusion coefficient $D_{\text{eff}}$ of the
protein along the membrane. The typical method to identify $D_{\text{eff}}$ is
to regard the temporal evolution of the mean squared displacement of the
protein using the relation
$\langle\Delta\mathbf{R}^2\rangle\equiv4D_{\text{eff}}^{\text{MSD}}t$, as
explaned above. Using the so determined value of the diffusion coefficient we
find for large times that the results in fig.~\ref{fig:two_time_regimes} are
well approximated by an exponential behaviour $\sim\exp[-t/\tau_D(k)]$ for
both $\gamma=1$ and $\gamma=6$. Overall, for $\gamma=1$ we observe that the
behaviour of the height correlations is well described by eq.~\eqref{eq:corr}.

\begin{figure}
  \centering
  \includegraphics[width=0.8\linewidth]{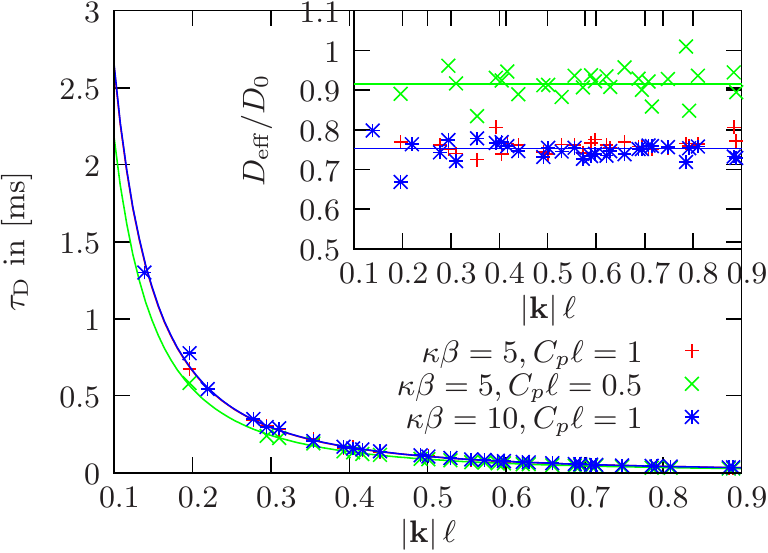}
  \caption{Diffusion dominated decay time $\tau_{\text{D}}(k)$ as a function
    of $k$ for $\gamma=1$ and the given values for $\beta\kappa$ and
    $C_p$. Symbols represent simulation results derived from fitting height
    correlation functions at later times; solid lines display the theoretical
    time scale using the effective diffusion coefficient determined through
    the mean squared displacement of the protein.
    In the inset the effective diffusion coefficient determined from the long
    time decay using $D_{\text{eff}}^{\text{M}}\equiv(k^2\tau_{\text{D}}(k))^{-1}$
    is plotted as a function of $k$ (symbols). The
    horizontal lines give $D_{\text{eff}}^{\text{MSD}}$ determined from the mean squared
    displacement.
  \label{fig:hh_long_tau}}
\end{figure}
\section{Diffusion coefficient extracted from membrane spectrum}
\label{sec:diff_spec}
\subsection{Determination of $D_{\text{eff}}$ from simulations}
In the discussion of fig.~\ref{fig:two_time_regimes} we used the mean squared
displacement to determine the diffusion coefficient of the protein. However,
the exponential decay of large time height correlations offers an alternative
method to extract $D_{\text{eff}}$.  The $\tau_D(k)$ resulting from
exponential fits to the late time decay as a function of $k$ are plotted in
the main graph of fig.~\ref{fig:hh_long_tau} for $\gamma=1$ but different
$\kappa$ and $C_p$. Using the previously determined value for the diffusion
coefficient we find a good agreement with $\tau_D(k)=(k^2D_{\text{eff}}^{\text{MSD}})^{-1}$
(solid lines). Thus without prior information on the mean squared displacement
of the proteins, it is possible to identify $D_{\text{eff}}$ from the height
correlations of the membrane using the behaviour of $\tau_D(k)$ as a function
of the wave-number $k$.  In the inset of fig.~\ref{fig:hh_long_tau} we plot
$D_{\text{eff}}^{\text{M}}\equiv(k^2\tau_D(k))^{-1}$ as a function of $k$. We find that the resulting
diffusion constant agrees very well with the diffusion constant $D_{\text{eff}}^{\text{MSD}}$.

\begin{figure}
  \centering
  \includegraphics[width=0.8\linewidth]{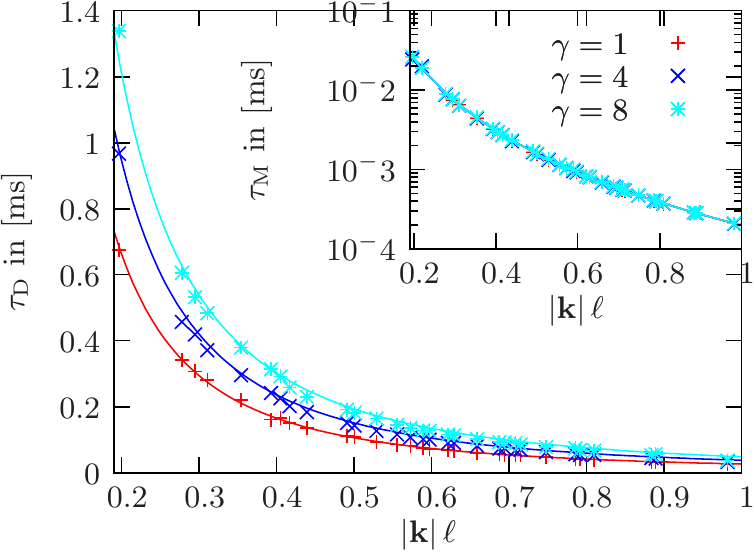}
   \caption{Diffusion time $\tau_{\text{D}}(k)$ as a function of $k$ for the
     given values of $\gamma$. All results apply for $C_p\ell=1$ and
     $\beta\kappa=5$. The symbols result from simulations while the solid
     lines are given by $(D_{\text{eff}}^{\text{MSD}}k^2)^{-1}$. The inset
     shows the corresponding membrane time scale $\tau_{\text{M}}(k)$
     determined by fitting the initial decay of the height correlation
     function (symbols) and given theoretically (solid line).
  \label{fig:long_time_decay_m}}
\end{figure}
While fig.~\ref{fig:hh_long_tau} only considers $\gamma=1$, we will now show
that the characteristic timescales $\tau_M$ and $\tau_D$ can also be
identified for $\gamma\neq1$.  The inset of fig.~\ref{fig:long_time_decay_m}
displays $\tau_M(k)$ as a function of $k$ as determined from the exponential
decay of the height correlations at short times for different values of
$\gamma$. Apart from $\gamma$ the other parameters of the membrane and the
particle are kept constant. We find that the results do not depend on the
rigidity of the protein. The dominant time scale for short
times is only determined by the properties of the membrane and coincides with
the correlation time of a freely fluctuating membrane without protein. In the
main plot the late time diffusive time scale $\tau_D$ is plotted as a function
of $k$. These results clearly depend on the rigidity of the protein, but only
because the diffusive motion is influenced by the protein-membrane interaction.
If we determine the effective diffusion coefficient from the mean
square displacement and compare the thus calculated $\tau_D$ (solid lines)
with the simulation results from the late time exponential fits to the
height correlations (symbols) the agreement is again
very good. Thus, if we had not had the possibility to determine the mean
squared displacement of the protein we could have determined $D_{\text{eff}}$
solely from the time dependence  of the height correlations.
\subsection{Estimate of experimental feasability}
\label{sec:discussion}  
\begin{figure}
  \centering
  \includegraphics[width=0.8\linewidth]{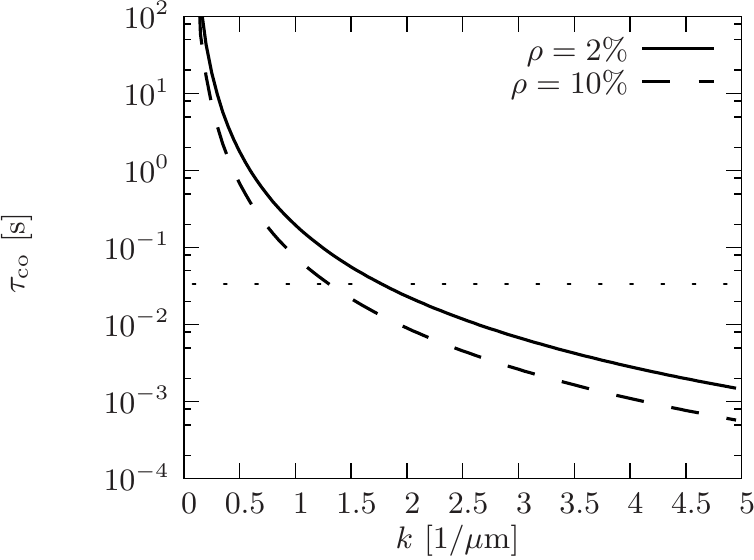}
  \caption{\label{fig:tau_co} Crossover time $\tau_{\text{co}}(k)$ from
    membrane fluctuation to diffusion dominated decay of height correlations
    as a function of wavenumber $k$, see eq.~\eqref{eq:tau_co}, for $\gamma=1$,
    $D_{\text{eff}}=10^{-8}$ cm$^2$/s, $\sqrt{\pi}a_pC_p=1$, and
    $\beta\kappa=10$. The dotted line corresponds to the typical resolution
    time of video microscopy.}  
\end{figure}
In the following we will show that our suggested method to determine the
lateral diffusion coefficient of proteins from height fluctuations in a
membrane is experimentally feasible. Height correlation functions can be
determined by video microscopy as explained in the introduction.  If
$\tau_{\text{co}}(k)$ is the crossover time from the decay of correlations
caused by the membrane dynamics to that dominated by the diffusive timescale
of the proteins, experiments must meet two conditions in order for the
crossover to become observable: on the one hand $\tau_{\text{co}}(k)$ must be
larger than the temporal resolution of the camera used in the
experiment. Values given in previous studies~\cite{Pecreaux:2004} are on the
order of $0.03$s. On the other hand $\tau_{\text{co}}(k)$ must be smaller than
the experimentally accessible total time that is on the order of minutes.
Since the crossover time is a function of the wave vector the restrictions for
$\tau_{\text{co}}(k)$ must, furthermore, be fulfilled for experimentally
accessible wave numbers. The spatial resolution of video microscopy allows for
$k$-values smaller than approximately $4\mu$m$^{-1}$.

If we have a general time-dependent function of the form $A\exp(-a
t)+B\exp(-b t)$ with $a\gg b$ a good estimate for the crossover time is given by
$(A+B)\exp(-a\tau_{\text{co}})=B\exp(-b\tau_{\text{co}})$. Within our
model the crossover time for $\gamma=1$ is given by
\begin{multline}
\tau_{\text{co}}(k)\equiv\left(\frac{1}{\tau_{\text{M}}(k)}-\frac{1}{\tau_{\text{D}}(k)}\right)^{-1}\times\\\times\ln\left[1+\left(\beta\kappa\,\rho\,\pi
    a_p^2C_p^2G(\mathbf{k})G(-\mathbf{k})\right)^{-1}\right]\,.
\label{eq:tau_co}
\end{multline}
While lateral diffusion coefficients of proteins in membranes are on the order
of $10^{-8}$cm$^2$/s~\cite{Gambin:2006}, the spontaneous curvature is not so well
determined.  In fig.~\ref{fig:tau_co}, we display the crossover time
$\tau_{\text{co}}(k)$ as a function of the wave number for different protein
densities $\rho$ and $D=10^{-8}$cm$^2$/s, $\beta\kappa=10$, and $\sqrt{\pi}
a_pC_p=1$. For the regarded protein densities we find that for small
wave numbers $\tau_{\text{co}}$ becomes larger than the typical temporal
resolution of experiments. We find that for wave numbers that lie within the
experimental range it should be possible to observe the two characteristic
time scales. Thus, the determination of the lateral diffusion coefficient from
the diffusion dominated decay of membrane height correlations should be
feasible. However, the interesting wave number range is reasonably narrow,
since $\tau_{\text{co}}$ is strongly increasing for smaller $k$.  Note, that
eq.~\eqref{eq:tau_co} is only valid for the situation of equal bending
rigidity of the protein and the membrane. In general this is obviously not the
case, however, the height correlations displayed in
fig.~\ref{fig:two_time_regimes} for $\gamma\neq1$ let us assume that the
crossover time is only weakly influenced by the bending rigidity of the
protein and that our estimate remains valid.
\section{Conclusions}

In this paper we have considered the influence of a protein interacting with a
fluctuating membrane via its bending rigidity and spontaneous curvature on
both the dynamics of the protein and the membrane. The quantities we have
looked at in detail are the lateral diffusion coefficient of the protein and
the height correlations of membrane fluctuations by use of
analytical calculations and Langevin simulations. We argue that the lateral
diffusion coefficient of a protein that interacts with the membrane is always
reduced compared to its  bare diffusion coefficient as long as there are no
external driving forces or active processes. Using a path integral approach we could derive an
analytical expression for this reduction that is valid within the lowest order
of a $(\beta\kappa)^{-1}$-expansion. Our simulations with parameters that
resemble those of real experiments show a wide applicability of this
expression. In addition a closer look at the correlation functions that
contribute to the reduction of the diffusion coefficient shows that the
correlations between the stochastic force acting on the protein and the
response of the membrane to the movement of the protein are responsible for
the reduction. 

The diffusion of the protein is obviously correlated with the height
correlations of the membrane.
The determination of the height correlations for the case of equal bending
rigidity of the protein and the membrane reveals that the
protein-membrane-interaction has a significant influence compared to a free
membrane. The most predominant feature is that the temporal decay of
correlations does not only display the timescale one would expect from the
membrane, but that the diffusive time scale of the influencing protein becomes
important. In realistic biomembrane systems these two time scales are well
separated such that a crossover from the initially fast decay of membrane
fluctuations to the slower protein diffusion dominated decay, is likely to be
observed in experiments. Since the decay at later times is directly related to
the effective diffusion coefficient of the protein, we suggest that the
measurement of membrane fluctuations might actually provide a means to
determine the lateral diffusion coefficient of the inserted proteins.

Our systematic approach to lateral diffusion of a protein interacting with the
shape of the membrane and the related influence on the membrane fluctuation
spectrum can be extended in various directions. The first question arising
from our study is to work out corrections to the
$(\beta\kappa)^{-1}$-expansion and to estimate their relevance. A further
perspective resulting from our analysis is the interesting limit
$\Lambda/\mu_0\rightarrow 0$ when the protein effectively moves in a fixed
membrane configuration. This situation is interesting theoretically, since the
diffusing particle no longer ``drags along'' the membrane, but is hindered in
its movement by potential barriers caused by the interaction of the particle
with the membrane.
Finally, while we have so far only considered membranes that are on average flat, the
extension to ruffled membranes poses an interesting challenge with significant
relevance for biological membranes like the endoplasmic reticulum or the
cristae in mitochondria.

\begin{acknowledgments}
We thank Thomas Speck for many helpful discussions.
\end{acknowledgments}


\begin{thebibliography}{42}
\expandafter\ifx\csname natexlab\endcsname\relax\def\natexlab#1{#1}\fi
\expandafter\ifx\csname bibnamefont\endcsname\relax
  \def\bibnamefont#1{#1}\fi
\expandafter\ifx\csname bibfnamefont\endcsname\relax
  \def\bibfnamefont#1{#1}\fi
\expandafter\ifx\csname citenamefont\endcsname\relax
  \def\citenamefont#1{#1}\fi
\expandafter\ifx\csname url\endcsname\relax
  \def\url#1{\texttt{#1}}\fi
\expandafter\ifx\csname urlprefix\endcsname\relax\def\urlprefix{URL }\fi
\providecommand{\bibinfo}[2]{#2}
\providecommand{\eprint}[2][]{\url{#2}}

\bibitem[{\citenamefont{Alberts et~al.}(1994))\citenamefont{Alberts, Bray,
  Lewis, Raff, Roberts, and Watson}}]{Alberts:1994}
\bibinfo{author}{\bibfnamefont{B.}~\bibnamefont{Alberts}},
  \bibinfo{author}{\bibfnamefont{D.}~\bibnamefont{Bray}},
  \bibinfo{author}{\bibfnamefont{J.}~\bibnamefont{Lewis}},
  \bibinfo{author}{\bibfnamefont{M.}~\bibnamefont{Raff}},
  \bibinfo{author}{\bibfnamefont{K.}~\bibnamefont{Roberts}}, \bibnamefont{and}
  \bibinfo{author}{\bibfnamefont{J.~D.} \bibnamefont{Watson}},
  \emph{\bibinfo{title}{Molecular Biology of the Cell}}
  (\bibinfo{publisher}{(Garland, NY}, \bibinfo{year}{1994)}).

\bibitem[{\citenamefont{Chiantia et~al.}(2009)\citenamefont{Chiantia, Ries, and
  Schwille}}]{Chiantia:2009}
\bibinfo{author}{\bibfnamefont{S.}~\bibnamefont{Chiantia}},
  \bibinfo{author}{\bibfnamefont{J.}~\bibnamefont{Ries}}, \bibnamefont{and}
  \bibinfo{author}{\bibfnamefont{P.}~\bibnamefont{Schwille}},
  \bibinfo{journal}{BBA-Biomembranes} \textbf{\bibinfo{volume}{1788}},
  \bibinfo{pages}{225} (\bibinfo{year}{2009}).

\bibitem[{\citenamefont{Lommerse et~al.}(2004)\citenamefont{Lommerse, Spaink,
  and Schmidt}}]{Lommerse:2004}
\bibinfo{author}{\bibfnamefont{P.~H.~M.} \bibnamefont{Lommerse}},
  \bibinfo{author}{\bibfnamefont{H.~P.} \bibnamefont{Spaink}},
  \bibnamefont{and} \bibinfo{author}{\bibfnamefont{T.}~\bibnamefont{Schmidt}},
  \bibinfo{journal}{BBA-Biomembranes} \textbf{\bibinfo{volume}{1664}},
  \bibinfo{pages}{119} (\bibinfo{year}{2004}).

\bibitem[{\citenamefont{Reits and Neefjes}(2001)}]{Reits:2001}
\bibinfo{author}{\bibfnamefont{E.~A.~J.} \bibnamefont{Reits}} \bibnamefont{and}
  \bibinfo{author}{\bibfnamefont{J.~J.} \bibnamefont{Neefjes}},
  \bibinfo{journal}{Nat.~Cell Biol.} \textbf{\bibinfo{volume}{3}},
  \bibinfo{pages}{E145} (\bibinfo{year}{2001}).

\bibitem[{\citenamefont{Cairo and Golan}(2008)}]{Cairo:2008}
\bibinfo{author}{\bibfnamefont{C.~W.} \bibnamefont{Cairo}} \bibnamefont{and}
  \bibinfo{author}{\bibfnamefont{D.~E.} \bibnamefont{Golan}},
  \bibinfo{journal}{Biopolymers} \textbf{\bibinfo{volume}{89}},
  \bibinfo{pages}{409} (\bibinfo{year}{2008}).

\bibitem[{\citenamefont{Lamb}(1959)}]{Lamb:1959}
\bibinfo{author}{\bibfnamefont{H.}~\bibnamefont{Lamb}},
  \emph{\bibinfo{title}{Hydrodynamics}} (\bibinfo{publisher}{Cambridge:
  University Press}, \bibinfo{year}{1959}).

\bibitem[{\citenamefont{Saffman and Delbr\"uck}(1975)}]{Saffman:1975}
\bibinfo{author}{\bibfnamefont{P.~G.} \bibnamefont{Saffman}} \bibnamefont{and}
  \bibinfo{author}{\bibfnamefont{M.}~\bibnamefont{Delbr\"uck}},
  \bibinfo{journal}{PNAS} \textbf{\bibinfo{volume}{{\bfseries 72}}},
  \bibinfo{pages}{3111} (\bibinfo{year}{1975}).

\bibitem[{\citenamefont{Peters and Cherry}(1982)}]{Peters:1982}
\bibinfo{author}{\bibfnamefont{R.}~\bibnamefont{Peters}} \bibnamefont{and}
  \bibinfo{author}{\bibfnamefont{R.~J.} \bibnamefont{Cherry}},
  \bibinfo{journal}{PNAS} \textbf{\bibinfo{volume}{79}}, \bibinfo{pages}{4317}
  (\bibinfo{year}{1982}).

\bibitem[{\citenamefont{Lee et~al.}(2003)\citenamefont{Lee, Revington, Dunn,
  and Petersen}}]{Lee:2003}
\bibinfo{author}{\bibfnamefont{C.~C.} \bibnamefont{Lee}},
  \bibinfo{author}{\bibfnamefont{M.}~\bibnamefont{Revington}},
  \bibinfo{author}{\bibfnamefont{S.~D.} \bibnamefont{Dunn}}, \bibnamefont{and}
  \bibinfo{author}{\bibfnamefont{N.~O.} \bibnamefont{Petersen}},
  \bibinfo{journal}{Biophys.~J.} \textbf{\bibinfo{volume}{84}},
  \bibinfo{pages}{1756} (\bibinfo{year}{2003}).

\bibitem[{\citenamefont{Cicuta et~al.}(2007)\citenamefont{Cicuta, Keller, and
  Veatch}}]{Cicuta:2007}
\bibinfo{author}{\bibfnamefont{P.}~\bibnamefont{Cicuta}},
  \bibinfo{author}{\bibfnamefont{S.~L.} \bibnamefont{Keller}},
  \bibnamefont{and} \bibinfo{author}{\bibfnamefont{S.~L.}
  \bibnamefont{Veatch}}, \bibinfo{journal}{J Phys Chem B}
  \textbf{\bibinfo{volume}{111}}, \bibinfo{pages}{3328} (\bibinfo{year}{2007}).

\bibitem[{\citenamefont{Gambin et~al.}(2006)\citenamefont{Gambin,
  Lopez-Esparza, Reffay, Sierecki, Gov, Genest, Hodges, and
  Urbach}}]{Gambin:2006}
\bibinfo{author}{\bibfnamefont{Y.}~\bibnamefont{Gambin}},
  \bibinfo{author}{\bibfnamefont{R.}~\bibnamefont{Lopez-Esparza}},
  \bibinfo{author}{\bibfnamefont{M.}~\bibnamefont{Reffay}},
  \bibinfo{author}{\bibfnamefont{E.}~\bibnamefont{Sierecki}},
  \bibinfo{author}{\bibfnamefont{N.~S.} \bibnamefont{Gov}},
  \bibinfo{author}{\bibfnamefont{M.}~\bibnamefont{Genest}},
  \bibinfo{author}{\bibfnamefont{R.~S.} \bibnamefont{Hodges}},
  \bibnamefont{and} \bibinfo{author}{\bibfnamefont{W.}~\bibnamefont{Urbach}},
  \bibinfo{journal}{PNAS} \textbf{\bibinfo{volume}{103}}, \bibinfo{pages}{2098}
  (\bibinfo{year}{2006}).

\bibitem[{\citenamefont{Guigas and Weiss}(2006)}]{Guigas:2006}
\bibinfo{author}{\bibfnamefont{G.}~\bibnamefont{Guigas}} \bibnamefont{and}
  \bibinfo{author}{\bibfnamefont{M.}~\bibnamefont{Weiss}},
  \bibinfo{journal}{Biophys.~J.} \textbf{\bibinfo{volume}{91}},
  \bibinfo{pages}{2393} (\bibinfo{year}{2006}).

\bibitem[{\citenamefont{Charitat et~al.}(2008)\citenamefont{Charitat, Lecuyer,
  and Fragneto}}]{Charitat:2008}
\bibinfo{author}{\bibfnamefont{T.}~\bibnamefont{Charitat}},
  \bibinfo{author}{\bibfnamefont{S.}~\bibnamefont{Lecuyer}}, \bibnamefont{and}
  \bibinfo{author}{\bibfnamefont{G.}~\bibnamefont{Fragneto}},
  \bibinfo{journal}{Biointerphases} \textbf{\bibinfo{volume}{3}},
  \bibinfo{pages}{FB3} (\bibinfo{year}{2008}).

\bibitem[{\citenamefont{Faucon et~al.}(1989)\citenamefont{Faucon, Mitov,
  M\'{e}l\'{e}ard, Bivas, and Bothorel}}]{Faucon:1989}
\bibinfo{author}{\bibfnamefont{J.~F.} \bibnamefont{Faucon}},
  \bibinfo{author}{\bibfnamefont{M.~D.} \bibnamefont{Mitov}},
  \bibinfo{author}{\bibfnamefont{P.}~\bibnamefont{M\'{e}l\'{e}ard}},
  \bibinfo{author}{\bibfnamefont{I.}~\bibnamefont{Bivas}}, \bibnamefont{and}
  \bibinfo{author}{\bibfnamefont{P.}~\bibnamefont{Bothorel}},
  \bibinfo{journal}{J.~Phys.~France} \textbf{\bibinfo{volume}{50}},
  \bibinfo{pages}{2389} (\bibinfo{year}{1989}).

\bibitem[{\citenamefont{Pecreaux et~al.}(2004)\citenamefont{Pecreaux,
  D\"obereiner, Prost, Joanny, and Bassereau}}]{Pecreaux:2004}
\bibinfo{author}{\bibfnamefont{J.}~\bibnamefont{Pecreaux}},
  \bibinfo{author}{\bibfnamefont{H.}~\bibnamefont{D\"obereiner}},
  \bibinfo{author}{\bibfnamefont{J.}~\bibnamefont{Prost}},
  \bibinfo{author}{\bibfnamefont{J.}~\bibnamefont{Joanny}}, \bibnamefont{and}
  \bibinfo{author}{\bibfnamefont{P.}~\bibnamefont{Bassereau}},
  \bibinfo{journal}{Eur.~Phys.~J.~E} \textbf{\bibinfo{volume}{13}},
  \bibinfo{pages}{277} (\bibinfo{year}{2004}).

\bibitem[{\citenamefont{\mbox{Rodr\'{\i}guez-Garc\'{\i}a}
  et~al.}(2009)\citenamefont{\mbox{Rodr\'{\i}guez-Garc\'{\i}a}, Arriaga, Mell,
  Moleiro, \mbox{L\'{o}pez-Montero}, and Monroy}}]{Rodriguez:2009}
\bibinfo{author}{\bibfnamefont{R.}~\bibnamefont{\mbox{Rodr\'{\i}guez-Garc\'{\i%
}a}}}, \bibinfo{author}{\bibfnamefont{L.~R.} \bibnamefont{Arriaga}},
  \bibinfo{author}{\bibfnamefont{M.}~\bibnamefont{Mell}},
  \bibinfo{author}{\bibfnamefont{L.~H.} \bibnamefont{Moleiro}},
  \bibinfo{author}{\bibfnamefont{I.}~\bibnamefont{\mbox{L\'{o}pez-Montero}}},
  \bibnamefont{and} \bibinfo{author}{\bibfnamefont{F.}~\bibnamefont{Monroy}},
  \bibinfo{journal}{Phys.~Rev.~Lett.} \textbf{\bibinfo{volume}{102}},
  \bibinfo{pages}{128101} (\bibinfo{year}{2009}).

\bibitem[{\citenamefont{Seifert and Langer}(1993)}]{Seifert:1993}
\bibinfo{author}{\bibfnamefont{U.}~\bibnamefont{Seifert}} \bibnamefont{and}
  \bibinfo{author}{\bibfnamefont{S.~A.} \bibnamefont{Langer}},
  \bibinfo{journal}{Europhys.~Lett.} \textbf{\bibinfo{volume}{23}},
  \bibinfo{pages}{71} (\bibinfo{year}{1993}).

\bibitem[{\citenamefont{Vitkova et~al.}(2006)\citenamefont{Vitkova,
  M\'{e}l\'{e}ard, Pott, and Bivas}}]{Vitkova:2006}
\bibinfo{author}{\bibfnamefont{V.}~\bibnamefont{Vitkova}},
  \bibinfo{author}{\bibfnamefont{P.}~\bibnamefont{M\'{e}l\'{e}ard}},
  \bibinfo{author}{\bibfnamefont{T.}~\bibnamefont{Pott}}, \bibnamefont{and}
  \bibinfo{author}{\bibfnamefont{I.}~\bibnamefont{Bivas}},
  \bibinfo{journal}{Eur.~Biophys.~J.} \textbf{\bibinfo{volume}{35}},
  \bibinfo{pages}{281} (\bibinfo{year}{2006}).

\bibitem[{\citenamefont{Girard et~al.}(2005)\citenamefont{Girard, Prost, and
  Bassereau}}]{Girard:2005}
\bibinfo{author}{\bibfnamefont{P.}~\bibnamefont{Girard}},
  \bibinfo{author}{\bibfnamefont{J.}~\bibnamefont{Prost}}, \bibnamefont{and}
  \bibinfo{author}{\bibfnamefont{P.}~\bibnamefont{Bassereau}},
  \bibinfo{journal}{Phys.~Rev.~Lett.} \textbf{\bibinfo{volume}{94}}
  (\bibinfo{year}{2005}).

\bibitem[{\citenamefont{\mbox{El Alaoui Faris}
  et~al.}(2009)\citenamefont{\mbox{El Alaoui Faris}, Lacoste, P\'ecr\'{e}aux,
  Joanny, Prost, and Bassereau}}]{Faris:2009}
\bibinfo{author}{\bibfnamefont{M.~D.} \bibnamefont{\mbox{El Alaoui Faris}}},
  \bibinfo{author}{\bibfnamefont{D.}~\bibnamefont{Lacoste}},
  \bibinfo{author}{\bibfnamefont{J.}~\bibnamefont{P\'ecr\'{e}aux}},
  \bibinfo{author}{\bibfnamefont{J.~F.} \bibnamefont{Joanny}},
  \bibinfo{author}{\bibfnamefont{J.}~\bibnamefont{Prost}}, \bibnamefont{and}
  \bibinfo{author}{\bibfnamefont{P.}~\bibnamefont{Bassereau}},
  \bibinfo{journal}{Physical Review Letters} \textbf{\bibinfo{volume}{102}}
  (\bibinfo{year}{2009}).

\bibitem[{\citenamefont{Seifert}(1997)}]{Seifert:1997}
\bibinfo{author}{\bibfnamefont{U.}~\bibnamefont{Seifert}},
  \bibinfo{journal}{Adv. Phys.} \textbf{\bibinfo{volume}{46}},
  \bibinfo{pages}{13} (\bibinfo{year}{1997}).

\bibitem[{\citenamefont{Goulian et~al.}(1993)\citenamefont{Goulian, Bruinsma,
  and Pincus}}]{Goulian:1993}
\bibinfo{author}{\bibfnamefont{M.}~\bibnamefont{Goulian}},
  \bibinfo{author}{\bibfnamefont{R.}~\bibnamefont{Bruinsma}}, \bibnamefont{and}
  \bibinfo{author}{\bibfnamefont{P.}~\bibnamefont{Pincus}},
  \bibinfo{journal}{Europhys.~Lett.} \textbf{\bibinfo{volume}{22}},
  \bibinfo{pages}{145} (\bibinfo{year}{1993}).

\bibitem[{\citenamefont{Netz and Pincus}(1995)}]{Netz:1995}
\bibinfo{author}{\bibfnamefont{R.~R.} \bibnamefont{Netz}} \bibnamefont{and}
  \bibinfo{author}{\bibfnamefont{P.}~\bibnamefont{Pincus}},
  \bibinfo{journal}{Phys.~Rev.~E} \textbf{\bibinfo{volume}{52}},
  \bibinfo{pages}{4114} (\bibinfo{year}{1995}).

\bibitem[{\citenamefont{Reister and Seifert}(2005)}]{Reister:2005}
\bibinfo{author}{\bibfnamefont{E.}~\bibnamefont{Reister}} \bibnamefont{and}
  \bibinfo{author}{\bibfnamefont{U.}~\bibnamefont{Seifert}},
  \bibinfo{journal}{Europhys.~Lett.} \textbf{\bibinfo{volume}{71}},
  \bibinfo{pages}{859} (\bibinfo{year}{2005}).

\bibitem[{\citenamefont{Reister-Gottfried
  et~al.}(2007)\citenamefont{Reister-Gottfried, Leitenberger, and
  Seifert}}]{Reister:2007}
\bibinfo{author}{\bibfnamefont{E.}~\bibnamefont{Reister-Gottfried}},
  \bibinfo{author}{\bibfnamefont{S.~M.} \bibnamefont{Leitenberger}},
  \bibnamefont{and} \bibinfo{author}{\bibfnamefont{U.}~\bibnamefont{Seifert}},
  \bibinfo{journal}{Phys.~Rev.~E} \textbf{\bibinfo{volume}{75}},
  \bibinfo{pages}{011908} (\bibinfo{year}{2007}).

\bibitem[{\citenamefont{Naji and Brown}(2007)}]{Naji:2007}
\bibinfo{author}{\bibfnamefont{A.}~\bibnamefont{Naji}} \bibnamefont{and}
  \bibinfo{author}{\bibfnamefont{F.~L.~H.} \bibnamefont{Brown}},
  \bibinfo{journal}{J.~Chem.~Phys.} \textbf{\bibinfo{volume}{126}}
  (\bibinfo{year}{2007}).

\bibitem[{\citenamefont{Leitenberger et~al.}(2008)\citenamefont{Leitenberger,
  Reister-Gottfried, and Seifert}}]{Leitenberger:2007a}
\bibinfo{author}{\bibfnamefont{S.~M.} \bibnamefont{Leitenberger}},
  \bibinfo{author}{\bibfnamefont{E.}~\bibnamefont{Reister-Gottfried}},
  \bibnamefont{and} \bibinfo{author}{\bibfnamefont{U.}~\bibnamefont{Seifert}},
  \bibinfo{journal}{Langmuir} \textbf{\bibinfo{volume}{24}},
  \bibinfo{pages}{1254} (\bibinfo{year}{2008}).

\bibitem[{\citenamefont{Gov}(2006)}]{Gov:2006}
\bibinfo{author}{\bibfnamefont{N.~S.} \bibnamefont{Gov}},
  \bibinfo{journal}{Phys.~Rev.~E} \textbf{\bibinfo{volume}{73}},
  \bibinfo{pages}{041918} (\bibinfo{year}{2006}).

\bibitem[{\citenamefont{Shlomovitz and Gov}(2008)}]{Shlomovitz:2008}
\bibinfo{author}{\bibfnamefont{R.}~\bibnamefont{Shlomovitz}} \bibnamefont{and}
  \bibinfo{author}{\bibfnamefont{N.~S.} \bibnamefont{Gov}},
  \bibinfo{journal}{Europhys.~Lett.} \textbf{\bibinfo{volume}{84}},
  \bibinfo{pages}{58008} (\bibinfo{year}{2008}).

\bibitem[{\citenamefont{Naji et~al.}(2009)\citenamefont{Naji, Atzberger, and
  Brown}}]{Naji:2009}
\bibinfo{author}{\bibfnamefont{A.}~\bibnamefont{Naji}},
  \bibinfo{author}{\bibfnamefont{P.~J.} \bibnamefont{Atzberger}},
  \bibnamefont{and} \bibinfo{author}{\bibfnamefont{F.~L.~H.}
  \bibnamefont{Brown}}, \bibinfo{journal}{Phys.~Rev.~Lett.}
  \textbf{\bibinfo{volume}{102}}, \bibinfo{pages}{138102}
  (\bibinfo{year}{2009}).

\bibitem[{\citenamefont{Leibler}(1986)}]{Leibler:1986}
\bibinfo{author}{\bibfnamefont{S.}~\bibnamefont{Leibler}},
  \bibinfo{journal}{Journal de Physique} \textbf{\bibinfo{volume}{47}},
  \bibinfo{pages}{507} (\bibinfo{year}{1986}).

\bibitem[{\citenamefont{Bivas and M\'el\'eard}(2003)}]{Bivas:2003}
\bibinfo{author}{\bibfnamefont{I.}~\bibnamefont{Bivas}} \bibnamefont{and}
  \bibinfo{author}{\bibfnamefont{P.}~\bibnamefont{M\'el\'eard}},
  \bibinfo{journal}{Phys. Rev. E} \textbf{\bibinfo{volume}{67}},
  \bibinfo{pages}{012901} (\bibinfo{year}{2003}).

\bibitem[{\citenamefont{Divet et~al.}(2002)\citenamefont{Divet, Biben, Cantat,
  Stephanou, Fourcade, and Misbah}}]{Divet:2002}
\bibinfo{author}{\bibfnamefont{F.}~\bibnamefont{Divet}},
  \bibinfo{author}{\bibfnamefont{T.}~\bibnamefont{Biben}},
  \bibinfo{author}{\bibfnamefont{I.}~\bibnamefont{Cantat}},
  \bibinfo{author}{\bibfnamefont{A.}~\bibnamefont{Stephanou}},
  \bibinfo{author}{\bibfnamefont{B.}~\bibnamefont{Fourcade}}, \bibnamefont{and}
  \bibinfo{author}{\bibfnamefont{C.}~\bibnamefont{Misbah}},
  \bibinfo{journal}{Europhys.~Lett.} \textbf{\bibinfo{volume}{60}},
  \bibinfo{pages}{795} (\bibinfo{year}{2002}).

\bibitem[{\citenamefont{Merath and Seifert}(2006)}]{Merath:2006}
\bibinfo{author}{\bibfnamefont{R.-J.} \bibnamefont{Merath}} \bibnamefont{and}
  \bibinfo{author}{\bibfnamefont{U.}~\bibnamefont{Seifert}},
  \bibinfo{journal}{Phys.~Rev.~E} \textbf{\bibinfo{volume}{73}},
  \bibinfo{pages}{010401R} (\bibinfo{year}{2006}).

\bibitem[{foo()}]{footnote}
\bibinfo{note}{The effect of the membrane curvature on the protein mobility
  derived from the projected flat trajectory in the $(x,y)$-plane is well
  described within a preaveraging approximation that integrates out the
  membrane fluctuations. The $\mu_{\text{p}}$ used in eq.~\eqref{eq:Langevin1}
  resembles the projected, preaveraged mobility related to the actual mobility
  $\mu\equiv D_{\text{0}}/k_\text{B}T$ along the membrane through
  $\mu_{\text{p}}/\mu=(1+\langle g^{-1}\rangle)/2$ with the metric
  $g\equiv1+(\partial_xh(\mathbf{r}))^2+(\partial_yh(\mathbf{r}))^2$~\cite{Rei%
ster:2007}.}

\bibitem[{\citenamefont{Lifson and Jackson}(1962)}]{Lifson:1962}
\bibinfo{author}{\bibfnamefont{S.}~\bibnamefont{Lifson}} \bibnamefont{and}
  \bibinfo{author}{\bibfnamefont{J.~L.} \bibnamefont{Jackson}},
  \bibinfo{journal}{J.~Chem.~Phys.} \textbf{\bibinfo{volume}{36}},
  \bibinfo{pages}{2410} (\bibinfo{year}{1962}).

\bibitem[{\citenamefont{Reimann et~al.}(2002)\citenamefont{Reimann, Van~den
  Broeck, Linke, H\"{a}nggi, Rubi, and P\'{e}rez-Madrid}}]{Reimann:2002}
\bibinfo{author}{\bibfnamefont{P.}~\bibnamefont{Reimann}},
  \bibinfo{author}{\bibfnamefont{C.}~\bibnamefont{Van~den Broeck}},
  \bibinfo{author}{\bibfnamefont{H.}~\bibnamefont{Linke}},
  \bibinfo{author}{\bibfnamefont{P.}~\bibnamefont{H\"{a}nggi}},
  \bibinfo{author}{\bibfnamefont{J.~M.} \bibnamefont{Rubi}}, \bibnamefont{and}
  \bibinfo{author}{\bibfnamefont{A.}~\bibnamefont{P\'{e}rez-Madrid}},
  \bibinfo{journal}{Phys.~Rev.~E} \textbf{\bibinfo{volume}{65}},
  \bibinfo{pages}{031104} (\bibinfo{year}{2002}).

\bibitem[{\citenamefont{Risken}(1996)}]{Risken:1996}
\bibinfo{author}{\bibfnamefont{H.}~\bibnamefont{Risken}},
  \emph{\bibinfo{title}{The Fokker-Planck Equation : Methods of Solutions and
  Applications (Springer Series in Synergetics)}}
  (\bibinfo{publisher}{Springer}, \bibinfo{year}{1996}).

\bibitem[{\citenamefont{Lin and Brown}(2004)}]{Lin:2004}
\bibinfo{author}{\bibfnamefont{L.~C.-L.} \bibnamefont{Lin}} \bibnamefont{and}
  \bibinfo{author}{\bibfnamefont{F.~L.~H.} \bibnamefont{Brown}},
  \bibinfo{journal}{Phys.~Rev.~Lett.} \textbf{\bibinfo{volume}{93}},
  \bibinfo{pages}{256001} (\bibinfo{year}{2004}).

\bibitem[{\citenamefont{Lin and Brown}(2005)}]{Lin:2005}
\bibinfo{author}{\bibfnamefont{L.~C.-L.} \bibnamefont{Lin}} \bibnamefont{and}
  \bibinfo{author}{\bibfnamefont{F.~L.~H.} \bibnamefont{Brown}},
  \bibinfo{journal}{Phys.~Rev.~E} \textbf{\bibinfo{volume}{72}},
  \bibinfo{pages}{011910} (\bibinfo{year}{2005}).

\bibitem[{\citenamefont{Brown}(2008)}]{Brown:2008}
\bibinfo{author}{\bibfnamefont{F.~L.} \bibnamefont{Brown}},
  \bibinfo{journal}{Ann.~Rev.~Phys.~Chem.} \textbf{\bibinfo{volume}{59}},
  \bibinfo{pages}{685} (\bibinfo{year}{2008}).

\bibitem[{\citenamefont{Frigo and Johnson}(2005)}]{FFTW05}
\bibinfo{author}{\bibfnamefont{M.}~\bibnamefont{Frigo}} \bibnamefont{and}
  \bibinfo{author}{\bibfnamefont{S.~G.} \bibnamefont{Johnson}},
  \bibinfo{journal}{Proceedings of the IEEE} \textbf{\bibinfo{volume}{93}},
  \bibinfo{pages}{216} (\bibinfo{year}{2005}).

\end{thebibliography}
\end{document}